\documentclass[reprint,amsmath,amssymb,rmp,onecolumn,notitlepage,11pt,numbers]{revtex4-1}
\setcitestyle{numbers} 
\setcitestyle{square} 
\usepackage[utf8]{inputenc}
\usepackage{graphicx}
\usepackage{float}
\usepackage{xcolor}
\usepackage{subcaption}
\usepackage[labelformat=parens,labelsep=quad,skip=3pt]{caption}
\usepackage{eurosym}

\usepackage{mathtools}

\newcommand{\coo}{\ensuremath{\mathrm{CO_2}} }
\usepackage{cleveref}
\usepackage{url}

\begin{document}

\title{Ride-pooling adoption model for \coo emission estimation}

\author{Milli Keil}
\affiliation{Technical University Berlin, Germany, Potsdam Institute for Climate Impact Research, Potsdam, Germany}
\author{Felix Creutzig}
\affiliation{Mercator Research Institute on Global Commons and Climate Change, Berlin, Germany, Technical University Berlin, Germany}
\author{Nora Molkenthin}
\affiliation{Potsdam Institute for Climate Impact Research, Potsdam, Germany}
\email{nora.molkenthin@pik-potsdam.de}

\begin{abstract}
With the climate emergency and growing challenges ranging from pollution to congestion, ride-pooling (rp) has been floated as a potential solution for less congested, low-carbon and more space-efficient urban transportation. However, it is unclear which system configurations will enable an economically viable case for shared pooled mobility. To develop a more profound comprehension of the mechanisms underlying this subject, we here develop a simplified model  to analyze the switching potential and \coo emissions of ride-pooling systems for a specified number of transport users, road network topology, and other system parameter values. This analysis is conducted across a broad range of switching probability functions (defined as the probability that a car or public transport user switches to ride-pooling) between an upper and lower bound of switching behaviour assumptions. Based on current Berlin parameters and the basic switching probability function, we find that ride-pooling can reduce the carbon emissions resulting from local transportation in Berlin by approximately 39\%. Policies that reduce the time factor - such as the provision of priority lanes - have the greatest effect in encouraging ride-pooling. For the system to be efficient and achieve measurable reductions in carbon emissions, the fleet size must be large enough. Across the range of switching probability functions, our results demonstrate that a fleet of 6,000 to 23,000 minibuses would be optimal to serve Berlin and reduce system-wide emissions.
\end{abstract}

\maketitle

\section{Introduction}

Passenger road transport is responsible for roughly 10\% of global \coo emissions \cite{sims2014transport, IEAemissionspassenger}. With the structural change from industrial to service-based economies, these emissions increase not only in absolute terms but also relative to other sectors \cite{creutzig2016evolving}. A classification of mitigation strategies includes the Avoid-Shift-Improve framework \cite{creutzig2018towards}: less demand for motorized transport through compact urban design (Avoid), a shift to public transit and cycling (Shift), and more efficient electric vehicles powered by renewable energy (Improve) \cite{skea2022summary}. While electric vehicles are central to any mitigation strategy, they will not reduce greenhouse gas (GHG) emissions quickly enough. In addition,
it would be desirable to reduce overall traffic and private car ownership for the added benefits of reducing noise, congestion, land use and accidents \cite{giesel2016impact}, as well as the overall energy needed. In recent years, the advancing digitalization has facilitated the rise of a sharing economy \cite{pouri2021digital}. Apart from car sharing, which disconnects car usage from car ownership, the sharing economy reveals options like dynamic ride-pooling, in which individual trips can be requested and pooled together, akin to shared taxi services. Models suggest that supplementing private motorized commuting with shared, pooled mobility in suburban and metropolitan areas could make a big difference, reducing GHG emissions by more than 30\% even without electrification \cite{itf2017shared}. Other studies estimate that shared mobility can reduce overall passenger transport emissions by at least 6.3\% \cite[p.9]{tikoudis2021exploring}. However, if we shift the focus from urban to rural ride-pooling, we see that it struggles to be economically viable and hardly generates the demand needed for significant pooling \cite{poltimae2022search,clewlow2017disruptive} - even though it is designed to fill gaps in public transport \cite{lygnerud2021business} and, according to \cite{manik2020topology}, is topologically more amenable to pooling.

Shared pooled mobility has long been studied as a generalization of the Traveling Salesperson Problem, called the  Dynamic Dial-A-Ride Problem (DARP)\cite{berbeglia2010dynamic}. DARPs can be directly addressed in agent-based models \cite{bischoff2018simulation, cich2017modeling}. Existing models also use different methodologies to analyze the demand for ride-pooling and its impact on, e.g., emissions and congestion. For example, \cite{kaufmann2023potential} uses statistical investigation to assess demand, \cite{storch2021incentive} and \cite{wolf2022spontaneous} develop a game-theoretic approach to explore demand under different ride-pooling incentives, and \cite{li2024determines} implements machine learning to calculate \coo emission reduction rates for ride-pooling.
However, since direct optimization quickly leads to impractically long computation times, the demand has to be downscaled \cite{kaddoura2021impact}. To cover the full demand, the problem is often solved approximately by using a set of heuristics \cite{herminghaus2019mean,alonso2017demand,engelhardt2020speed,manik2020topology}. However, even these heuristic simulations incur significant computational costs, limiting the feasibility of exploratory, empirical studies. 

Therefore, we introduce a simplified model that does not claim to provide an exact quantitative outcome for policy, but rather approximates the possible effects of various interventions on (economic and transportation) decisions for the ride-pooling market. 
In this work, the switching probability is defined as the probability that an individual will switch their former mode of transport (either car or public transport) to ride-pooling. Thus, we construct a simple function that includes important, averaged parameter values yet remains analytically tractable. This model is based on the scaling behaviour of on-demand ride-pooling, together with a set of very general constraints on the shape of the switching probability. While this does not allow for concrete predictions, we find that the order of magnitude of the approximated fleet size is consistent with concrete simulations for the city of Berlin \cite{schmaus2023shared}. This allows us, for the first time, to estimate the ride-pooling fleet size without knowing the pooling algorithm or the adoption probability. A numerical comparison of several alternative switching probability functions, including upper and a lower bounds, reveals interesting mode choice tendencies and estimates effects of parameter modifications. The question is thus related to the so-called minimum fleet problem, which is predicated on computing the optimal number of buses required to accommodate the maximum number of requests, while ensuring that all vehicles are fully utilized \cite{muhle2023analytical}. 

Our simplified approach is intended to provide a broad overview of the potential effects of an additional ride-pooling service on equilibrium mode choice.
The approximation is based on analytical results of on-demand ride-pooling derived in \cite{molkenthin2020scaling} for average travel times and vehicle occupancies in the limit of large fleet sizes and request rates. While the model works with unbounded vehicle sizes, it has been shown in \cite{zech2022collective} to be a valid approximation as long as the typical occupancy is sufficiently smaller than the vehicle size (about 50 to 100\% larger vehicle sizes). The mode choice potential is modeled as a \emph{switching probability function}, which estimates the average fraction of passengers changing their original mode of transport (car or public transport) based on average relative prices and travel times of pooled rides compared to their original modes of transportation. Since we only model approximate fractions, rather than individual behaviour, individual preferences and influences, such as feelings of security or comfort are disregarded in this study.

We then use this simplified model to estimate the optimal size and number of ride-pooling vehicles for given scenarios. However, due to the simplicity of the model, the results do not constitute realistic predictions for mode choices, but rather roughly map out possible effects on a large space of mode choice potentials.

\section{Methods: Model characteristics}

The model considers three modes of transportation: private car, public transport, and ride-pooling while omitting all other forms of passenger transportation (e.g., cycling, walking, taxi driving). Thus, we only consider non-walkable/cyclable trips, and ignore taxi trips as too small a fraction to have much impact. In addition, only the potential to switch from car to ride-pooling or from public transport to ride-pooling is taken into account, while the choice between car and public transport is considered to be already fixed (i.e. due to car ownership or annual season ticket). We think of our model as representing a transport system before and after the introduction of a ride-pooling service, while not changing price or speed of any other mode. Therefore, the motivation to switch between car and public transport would come from secondary effects such as changes in road traffic, which are likely to be small. We also assume that total demand is constant and does not change with the addition of the pooling service. After the introduction of the pooled service, we assume that enough time has passed for the system to reach a new equilibrium, i.e., not only that the information about the service has reached everyone, but also that everyone has responded to the changes in price and travel time (any other influences, such as privacy or convenience, are not considered).

We assume that car users change their mode of transport to ride-pooling at a switching probability of $P_{car}$ and public transport users change their mode of transport with the probability of $P_{pt}$, where the subscript \emph{car} stands for a variable associated with car users, \emph{pt} for variables associated with public transportation, and \emph{rp} for variables associated with ride-pooling. A comprehensive list of all variables, accompanied by their precise definitions, can be found in Table (I) of the Supplemental Information. The switching probabilities define the load $\lambda$ of the ride-pooling system as the number of ride-pooling users per unit of time $\tau$, which is the sum of switched car drivers and switched public transport riders (and thus the total demand for ride-pooling). 

\begin{equation}
    \lambda=(M_{car} P_{car}+M_{pt} P_{pt}) \lambda_{tot}  
    \label{eq.lambda_rs}
\end{equation}

Here $M_{car}$ is the initial proportion of car drivers before the introduction of the ride-pooling mode, $M_{pt}$ is the initial proportion of public transport users and $\lambda_{tot}$ is the total number of trip requests for cars and public transport on the system within our natural timescale $\tau$. 
The total travel demand $\lambda_{tot}$ is estimated from the population of the area, using average travel habits recorded in the literature and is estimated to be: 
\begin{equation}
    \lambda_{tot}=\frac{3.2 \text{ trips}}{24 \text{ hours}} \; \tau \; d \; , \;\;\;\; \tau = \frac{l}{v} 
\end{equation}
Where the natural timescale $\tau$ of the problem is the average time it takes to drive the requested trip on a direct route of average length $l$ km with average speed $v$ km/h. $d$ is the number of inhabitants in the system with 3.2 trips/day/person resulting from MiD data \cite[table 3]{nobis2018mobilitat}.

The switching probability function estimates the average fraction of passengers switching from one mode of transport to another. The probability of switching from a previous mode is denoted by $P_{ref}$, where the reference (\textit{ref}) can be either car or public transport - the function is further elaborated in Eq. \ref{eq.sr}. This should be conceptualized as an aggregate quantity that provides a probability of switching averaged over the population. While individual agents value time, money, and other factors heterogeneously, we model only the fraction of agents that decide to switch at a given relative travel time and relative price. 
The total duration of the ride-pooling trip, referred to as service time $t_s$, is the sum of the waiting time for a ride-pooling vehicle, and the subsequent travel time to the designated stop (further elaborated in Eq. \ref{eq.ts}). $C_B$ on the other hand defines the price per passenger for the pooled ride (as shown in Eq. \ref{eq.Cb}).
For the functional form of the switching probability functions $P_{ref}$, we consider the following constraints concerning the service time $t_s$ and cost $C_B$ for the ride-pooling alternative:
\begin{enumerate}
    \item[i)] If $t_s$ or $C_B \rightarrow \infty$ then $P_{ref} \rightarrow 0$
    \item [ii)] If $t_s$ and $C_B \rightarrow 0$ then $P_{ref} \rightarrow \alpha$
    \item[iii)] If $t_s=t_{ref}$ and $C_B=C_{ref}$ then $\beta_1 < P_{ref} < \beta_2$
    \item[iv)] If either temporal \textbf{or} financial values vary, the function reacts monotonously
\end{enumerate}
Here, $\alpha$, $\beta_1$, and $\beta_2$ are all numbers between 0 and 1.
Requirement i) reflects the assumption that no agent would switch to a pooling service that takes infinitely long or is infinitely expensive. ii) ensures that if a free and immediate service were to emerge (e.g. teleportation), it would eventually convince all but the fraction $\alpha$ who, for whatever reason, cannot / will not use ride-pooling.
iii) constrains the probability of switching for identical price and service quality to be between $\beta_1$ and $\beta_2$. This is to make sure that with identical service quality (to the user's current mode, which takes service time $t_{ref}$ and costs $C_{ref}$), a certain fraction is willing to share their trip.
iv) ensures that an increase in price with fixed travel time or an increase in travel time with fixed price, monotonically reduces the probability of switching.

As the true values for the maximal pooling fraction $\alpha$ and the pooling fraction at equal service $\beta$ are unknown and there is no suitable study estimating them, we will assume that everyone can be swayed by zero cost and instant arrival and set $\alpha=1$ (although instantaneous arrival is obviously not physical). At the same level of service (times and prices are the same), we assume that no more than $\beta_2=1/2$ of the population would accept the new ride-pooling option (equivalent to a random choice), but at least $\beta_1=1/4$ would (arbitrary choice). Choosing $\beta_1$ and $\beta_2$ puts the ride-pooling option at a slight disadvantage, but limits the reluctance to change, modeling a bias towards habitual transport modes.
Out of the numerous functions fulfilling these criteria, we select a particularly simple one to carry out the  main analytical analysis of its effects on a wide range of parameters (Section III). Subsequently, we show in a numerical analysis of other functions, including the logit function (which is widely used in economics \cite{quinet2004principles}) as well as functions representing upper and lower bounds for the switching probabilities $P_{ref}$, that the results remain qualitatively valid (Section IV). Visualizations of all switching probability functions can be found in Fig. 2 of the Supplemental Information.

Based on these observations, we expect the behaviour of the money and time dependencies to fall between an upper and lower bound (or the probability functions with the highest/lowest switching probabilities compatible with constraints i)-iv)), regardless of the exact functional form. Our function takes into account essential factors for the transportation decision of the consumers (financial and temporal ones). However, the functions do not claim to be quantitatively exact and do not consider other psychological factors, such as personal space or safety concerns, that also play a role in decision making. 

We now construct a simple switching probability function based on the above requirements i), ii), iii), and iv), which allows for an analytical solution. 
It is the product of the inverse relative time and price:
\begin{align}
    P_{ref}&=\text{Pr(choosing ride-pooling over reference})=\frac{1}{\frac{t_s}{t_{ref}}+1}\frac{1}{\frac{C_B}{C_{ref}}+1} 
    \label{eq.sr}
\end{align}
This function approaches $0$  when at least one of $t_s$ and $C_B$ becomes infinite. It approaches $1$ if both $t_s$ and $C_B$ reach 0. Note, however, that it will take the value $\beta_1$ ($1/4$) if both travel time and cost are equal to that of a person's previous mode, which is the lower end of the range in iii).
Numerical analysis of other switching probabilities can be seen in Fig.~\ref{fig.alternative_sr}.

The average car and public transport travel times $t_{car}$ and $t_{pt}$, as well as car and public transport costs $C_{car}$ and $C_{pt}$ are derived from a range of parameters, such as approximate average prices of transport tickets ($C_{pt}$) and approximate times of searching for a parking lot ($t_p$), which we assume to be fixed system parameters as shown in Table \ref{tab:value definitions}. The parameters are set as fixed system parameters for the Base Case; however, the effects of changing individual parameters are explored in section III. The remaining two parameters are the average service time $t_s$ and the average ride-pooling price $C_B$. Both are approximated based on the perfect service limit found in \cite{molkenthin2020scaling, zech2022collective}, which is the limit of many requests and many vehicles. The average driving time $t_d$ quickly reaches the average direct travel time $\tau$ (used as time unit) and the average waiting time $t_w$ approaches zero with $1/B$, where $B$ is the fleet size, or number of ride-pooling vehicles [buses] operating in the system. This gives:
\begin{equation}
    t_s=t_w+t_d\approx \gamma \tau B^{-1}+\tau
    \label{eq.ts}
\end{equation}
where $\gamma$ is a parameter quantifying the effect of road network topology on the efficiency of ride-pooling for the region in question, as defined and estimated in \cite{molkenthin2020scaling}.

The average price $C_B$ per ride-pooling user is calculated by dividing the operating expenses of one ride-pooling vehicle by the number of passengers sharing it. Thus, the price $C_B$ for a pooled ride is approximated as the price $C$ for operating the vehicle divided by the average number $O$ of occupants in the vehicle. While $C$ is a constant parameter (computed as the sum of fuel and driver costs), estimated from literature, $O$ is estimated from the perfect service limit in \cite{molkenthin2020scaling}. As a result we get:
\begin{align}
    C_B &=\frac{C}{O} \approx \frac{C B}{\lambda} 
    \label{eq.Cb}
\end{align}
By inserting Eq.~\ref{eq.Cb} into Eq.~\ref{eq.sr}, which is then put into Eq.~\ref{eq.lambda_rs}, we make all dependence on $\lambda$ explicit. Thus, we obtain a consistency equation, is solved for $\lambda$ to express it as a function of $B$ (see the Supplemental Information (SI) Eq.~1-5 for details).
\begin{align}
    \lambda(B) &=-\frac{CB}{2}(\frac{1}{C_{car}}+\frac{1}{ C_{pt}})+\frac{\lambda_{tot}}{2} ( \frac{M_{car}t_{car}}{t_s+t_{car}}+ \frac{M_{pt}t_{pt}}{t_s+t_{pt}}) \\
    &+\sqrt{\frac{C^2B^2}{4}(\frac{1}{C_{car}}-\frac{1}{C_{pt}})^2+\frac{\lambda_{tot}^2}{4}(\frac{M_{car}t_{car}}{t_s+t_{car}}+\frac{M_{pt}t_{pt}}{t_s+t_{pt}})^2+\frac{CB\lambda_{tot}}{2}(-\frac{M_{car}t_{car}}{t_s+t_{car}}+\frac{M_{pt}t_{pt}}{t_s+t_{pt}})(\frac{1}{C_{car}}-\frac{1}{C_{pt}})}\nonumber
    \label{eq.lambda}
\end{align}
While the equation has three roots, we select the one convincing the largest fraction $\lambda(B)$ to use ride-pooling.

This function for $\lambda(B)$ can now be used to evaluate the switching probabilities for different values of $B$ and thus gives the modal split after the introduction of ride-pooling.

We use this modal split together with estimates for the \coo footprint per time unit $\tau$ of the different modes to approximate the system-wide \coo emissions after the introduction of ride-pooling for a set of parameters. Note that emissions per unit $\tau$ are equivalent to emissions per direct travel distance as all vehicles drive at the constant velocity $v=l/\tau$.
The \coo footprint for private cars in the system, denoted by $CO_2^{car}$ is given by the direct distance $l$ of the trip, plus the distance $l_p$ driven to find a parking lot multiplied by the number of trips taken by car and the average emissions $E_{car}$ per car per distance:
\begin{equation}
    CO_2^{car}= E_{car}(l+l_p)(M_{car}(1-P_{car}(B))\lambda_{tot})
\end{equation}
The emissions of the ride-pooling service $CO_2^{rp}$ per $\tau$ are given by the emissions per ride-pooling vehicle $E_{rp}$ multiplied by the number of vehicles $B$ and the direct distance of the requested trip $l$. This is because in the assumed perfect service limit, all $B$ ride-pooling vehicles always travel at velocity $v$, since the probability of vehicles being empty is extremely low in this limit. Thus, detours and occupancy have no impact on the \coo footprint of the service. Regardless of occupancy, we use one vehicle size throughout the analysis of fleet sizes $B$. This leads to an overestimation of the impact of ride-pooling for small fleet sizes $B$, as the occupancy may be unrealistically high and thus generate high \coo savings (but also an underestimation for large fleet sizes $B$). However, since the emissions per vehicle scale sub-linearly with the number of seats, we do not expect this approximation to change the general results.
\begin{equation}
    CO_2^{rp}= E_{rp} l B
\end{equation}
Finally, the \coo footprint of the public transport system $CO_2^{pt}$ per $\tau$ is assumed to be constant and also independent of the number of users, as its changes operate on different time scales compared to ride-pooling and private car. In the context of the present analysis, only the short-term implications and effects of the introduction of ride-pooling are considered.

The \coo emissions of the public transport services for the city of Berlin are given in \cite[p.100]{nachhaltigkeitsbericht2020} as 163,000 tonnes per year. This number excludes trams, metros, and electric buses, since they run on renewable electricity, but includes emissions from heating the company's office and service buildings. As such, it is likely to slightly overestimate bus emissions relative to the other modes, which are based entirely on emissions from vehicle miles traveled.
\begin{equation}
    CO_2^{pt}= (163,000 \; \frac{\text{t}}{\text{year}}/\; 365\;  \frac{\text{days}}{\text{year}}/\; 24\;  \frac{\text{h}}{\text{day}}) \text{ } \tau = 11.75 \; \text{t}
\end{equation}

In the long term, there would be an adjustment to the public transport supply if the overall demand were to be altered; however, this will not be evaluated in the present work. In this sense the estimation is an upper bound on the emissions, making sure that gains are not made at the expense of line service quality.
Taken together, we thus get the following approximation for the system-wide \coo emissions $CO_2^{tot}$ per time $\tau$:
\begin{equation}
    CO_2^{tot}(B)=CO_2^{car}+CO_2^{rp}+CO_2^{pt}=E_{car}(l+l_p)(M_{car}(1-P_{car}(B))\lambda_{tot})+E_{rp}lB+CO_2^{pt}
    \label{eq.coo}
\end{equation} 

\section{Results}
\subsection{Parameters Influencing Mode Selection Potential}
In the upcoming section, we analyze pooled mobility demand $\lambda$, system-wide \coo emissions $CO_2^{tot}$ and vehicle occupancy $O$ as functions of the ride-pooling fleet size $B$ for a range of parameter values. The reference parameter choices are based on literature values for the city of Berlin and given in Table \ref{tab:value definitions}.

\begin{table}[b]
 \begin{center}
 \caption{Parameter values for the urban area of Berlin (with $d = 3.7 \times 10^6$ inhabitants). All abbreviations used are defined in Table (I) in the Supplementary Information.}
    \label{tab:value definitions}
    \begin{tabular}{l|l|l}
      \textbf{parameter} & \textbf{value} & \textbf{references}\\
      \hline
      $\lambda_{tot}$ - trip requests per time unit $\tau$       &   $\lambda_{tot} = \frac{3.2}{24\;\text{h}} \; \tau \; d = 310,800$&  \cite[table 3]{nobis2018mobilitat} \\
    $M_{car}$ - initial proportion of car      &   0.38       &   \cite[p. 47]{MiDErgebnisbericht2018}\\
    $M_{pt}$ - initial proportion of public transport      &   0.2       &   \cite[p. 47]{MiDErgebnisbericht2018}\\
    $\tau$ - natural time scale                            & $\frac{l}{v} = \frac{12 \; \text{km}}{19 \; \text{km/h}} = 0.63$ h    & $l$: \cite[Tab. 3]{MiDErgebnisbericht2018}, $v$: \cite{tachet2017scaling} \\
    $t_{car}$ - average trip time using car & $\tau + t_p$ $= \tau + 0.125 \; \text{h} = 0.755$ h & average from \cite{Statista2017Parken}   \\
    $t_{pt}$ - average trip time public transport & $\epsilon \; \tau = 2 \; \tau = 1.26 $ h&  \cite{liao2020disparities} \\
    $C_{op}$ - car operating costs per km & $0.0868 \;\frac{\mbox{\euro}}{\text{km}}$ & \cite{BenzinPreiseStatista}\\
    $C_{car}$ - cost per car trip& $C_{op} \;l+C_p = C_{op} \; l+6 \; \mbox{\euro}=7.04$ \euro & $C_p$:\cite{Statista2017Parkkosten}\\
    $C_{pt}$ - cost public transport & 3.6 \euro &  \cite{VBB}  \\
      $\gamma$ - topological efficiency    & Berlin: 585        &  \cite[suppl.]{molkenthin2020scaling}    \\
      $l_p$ - distance for finding a parking lot    &   $t_p\; v_p = \frac{7.5}{60}\;\text{h }10\;\frac{\text{km}}{\text{h}}=1.25 $ km      &    time$\times$velocity (estimated)   \\
       $C$ - cost for one ride-pooling bus per unit time   &    $0.3\;\frac{\mbox{\euro}}{\text{km}}\;l+10\;\frac{\mbox{\euro}}{\text{h}}\;\tau=9.9$ \euro     &   fuel+driver (estimated)  \\
      $E^{car}$ - emissions per km car         &   $185.5$ g/km       &  \cite[p. 309]{BMVI2020Zahlen}      \\
      $E^{rp}$ - emissions per km ride-pooling        &   $736.7$ g/km       &  \cite[Tab. 6]{schmied2014berechnung}\\ 
    \end{tabular}
\end{center} 
\end{table}
The number of trips requested $\lambda_{tot}$ per time $\tau$ across all modes of transport is calculated based on the number of inhabitants $d= 3.7 \times 10^6$ with an average amount of requests of 3.2 trips per day per person \cite[table 3]{nobis2018mobilitat} resulting in $\lambda_{tot}=310,800$ trips per time $\tau$. In order to analyze different settings and their respective \coo emissions, we vary $\lambda_{tot}$ within an interesting range to compare the travel demand of medium cities with that of megacities. We scale $\lambda_{tot}$ down to $10^5$, which is comparable to a city of 1.2 million people and also up to $5\times10^5$, which could be the case for a city twice the size of Berlin, such as New York. Of these requested trips, a share of $M_{car}=0.38$ \cite[p. 47]{MiDErgebnisbericht2018} is made by car and a share of $M_{pt}=0.2$ \cite[p. 47]{MiDErgebnisbericht2018} is made by public transport. Most of the other trips are made by active modes (walking, cycling) and are not considered in the further analysis. The modal share for private cars, $M_{car}$ is varied between $0.28$ and $0.48$, with $M_{pt}$ varying, assuming a fixed share of motorized individual transport ($M_{pt} = 0.58-M_{car}$). This variation is based on the observation that other types of areas (not metropolitan areas) such as smaller cities or rural areas, have less public transportation availability \cite[p. 47]{MiDErgebnisbericht2018}. However, some bigger cities may even have a lower proportion of car drivers. The natural time scale $\tau= 0.63$ h is the average direct driving time and is calculated based on the average requested trip distance $l$ divided by the average speed $v$ as reported in \cite{MiDErgebnisbericht2018,tachet2017scaling}. 

In the following, we do not vary $\tau$ directly, but vary the ride-pooling speed $v_{rp}$ (e.g., when express lanes are allowed to be used by ride-pooling vehicles or when high occupancy vehicle (HOV) lanes are introduced). This results in a distinct ride-pooling timescale $\tau_{rp}$. 
The average private car travel time $t_{car}$ is calculated by adding the average time to find a parking space $t_p$, as found in \cite{Statista2017Parken} (plus some extra time to walk to the destination), to the average direct travel time $\tau$. In the parameter modification section, we analyze the effect of increased search time for a parking space $t_p$, as this can be influenced directly by policy interventions (removing parking spaces and thus increasing the time spent searching for a parking space). $t_{pt}$ as public transport time is calculated using a factor $\epsilon = 2 $ \cite{liao2020disparities} multiplied by the average direct travel time. This multiplier will be varied throughout the paper using the range of 1.4 to 2.6 \cite{liao2020disparities}.
Regarding the costs of owning and operating a personal car $C_{car}$, we consider a compact car from Volkswagen (Passat Variant 1.5 with approximately 5.5 l / 100 km fuel consumption) to be representative because these are among the most commonly owned types of cars in Germany \cite{Pressemitteilung2021}. Note that the cost of operating a car is calculated on a per-kilometer basis (in economics known as "marginal cost per kilometer") because this is a metric that has been shown to be most frequently considered by car drivers, as opposed to fixed costs, which include purchasing and maintaining a vehicle \cite{andor2020running}. During the parameter change, $C_{op}$, the operating cost per kilometer for a private car, is varied so that either the cost of driving a certain distance $l$ is perceived as being zero (which could be the case with company cars), or it is increased so that fixed costs (maintenance, purchase) are included in the price per kilometer. The cost for parking a private car, $C_p$ is varied in the same way as the time to find a parking space $t_p$ - assuming either policy interventions with high parking prices ($C_p = 15$ \euro) or no parking costs ($C_p = 0$ \euro).
The price for a single public transport ticket $C_{pt}$ is set to $3.6$ \euro. Variations include lowering the price to $1$ \euro \space and $0.5$ \euro. The topological factor $\gamma$ is used as defined and calculated in \cite{molkenthin2020scaling}. The value depends on the network topology: while the smaller city Göttingen has a more centralized topology and thus a larger topological factor ($\gamma=996$), making it harder to share rides, the rural area of Ostfriesland has a small topological factor $\gamma=213$, owing to the mesh-like rural street network. 
For $C$, the cost for one ride-pooling bus per unit time $\tau$, fixed costs are neglected. This means that only fuel and driver costs are included, estimated on the basis of the current minimum wage and the fuel consumption of an average city bus. $l_p$ is the distance traveled to find a parking lot and plays a role in the \coo emissions calculation. To calculate the total emissions in the system $CO_2^{tot}$, we also need to take into account the additional distance traveled by car to find a parking space. The values are estimated based on the average parking search times $t_p$ for Berlin found in \cite{Statista2017Parken} and the approximate speed driven during the search $v_p$. $E_{car}$ and $E_{rp}$ represent the emissions per kilometer for car and ride-pooling. 
For the \coo emissions of an average ride-pooling bus $E_{rp}$, we assume that one liter of diesel fuel emits on average 2650 g \coo \cite{buchal2019kohlemotoren}. In addition, we assume that our buses are minibuses or midibuses (a bus that is lighter than 15 tons and has about 22 seats) and we know from \cite[tab.6]{schmied2014berechnung} that these buses use on average $27.8$ liters of diesel per 100 kilometers, leading to $736.7$ g/km for $E_{rp}$.

\begin{figure}
    \centering
    \includegraphics[width= \columnwidth]{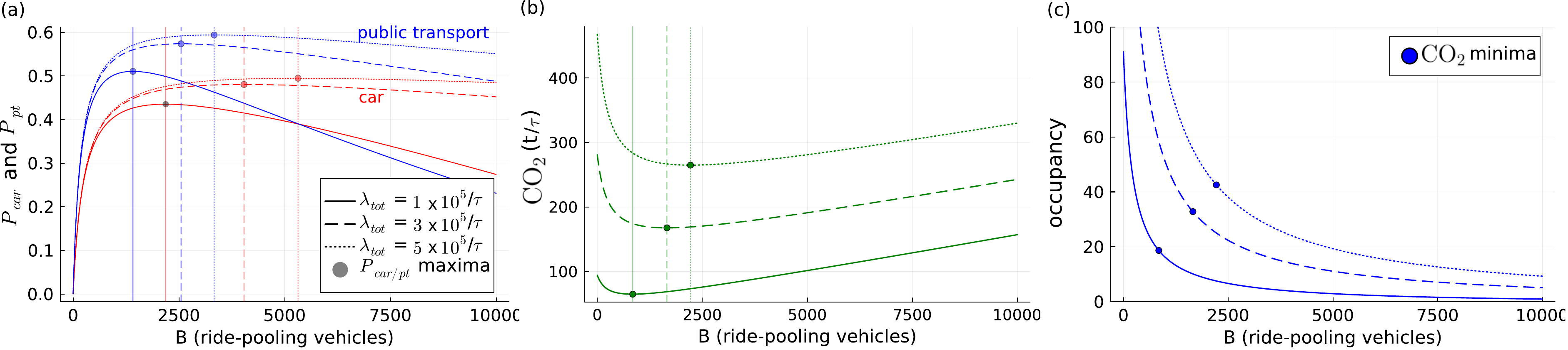}
    \caption{Ride-pooling switching probabilities $P_{car}$ and $P_{pt}$, the resulting \coo emissions and corresponding vehicle occupancies with fleet size $B$ for different total number of requests $\lambda_{tot}$ per time unit $\tau$. a) Ride-pooling adoption increases as the service improves and decreases again as the costs rise. b) \coo emissions per time unit $\tau$ decrease as adoption rates increase and increase again as vehicle occupancy decreases. c) Vehicle occupancy decreases with the number of vehicles.}
    \label{fig:SandCO2}
\end{figure}

The resulting switching probabilities for private car drivers and public transport users with the Berlin parameters from Tab.~\ref{tab:value definitions} and variations of the demand per unit time $\lambda_{tot}$ over different fleet sizes $B$ are shown in Fig.~\ref{fig:SandCO2}(a). Fig.~\ref{fig:SandCO2}(b) shows the \coo emissions for the same three demand scenarios and Fig.~\ref{fig:SandCO2}(c) shows the corresponding vehicle occupancies as a function of fleet size $B$.

All \coo curves shown in Fig.~\ref{fig:SandCO2}(b) are qualitatively similar. Starting from fleet size $B$ equal to zero, they rapidly decrease as the number of shared vehicles increases. The curves reach a minimum, after which the trend reverses and the function slowly rises. The initial reduction in emissions is due to the increasing switching probability (shown in Fig.~\ref{fig:SandCO2}(a)) as average ride-pooling service times decrease and the price per ride-pooling user is still low due to the small number of ride-pooling vehicles in the system. Eventually, however, the increase in emissions from ride-pooling vehicles outweighs the reduction in emissions from private cars, causing \coo emissions to slowly increase again and at one point exceed the original \coo value (due to the larger size of ride-pooling vehicles compared to private cars, as well as deadheading). As occupancy decreases, deadheading becomes increasingly prevalent, potentially resulting in an increase in the average distance traveled per journey.

However, since we consider our vehicles to be traveling continuously, the \coo function grows indefinitely and becomes invalid when the number of vehicles $B$ exceeds the number of requested trips per unit time $\lambda_{tot}$. The average vehicle occupancy shown in Fig.~\ref{fig:SandCO2}(c) decreases as the fleet size $B$ increases. Note that for small fleet sizes, the occupancies are well above the typical vehicle sizes used in on-demand pooling systems. The \coo minima are marked with dots in the figure, indicating that at the emissions optimum the average vehicle occupancy is between 20 and 40, depending on the system load $\lambda_{tot}$. Considering that a real system would have to account for fluctuations, we estimate based on the results of \cite{zech2022collective} that capacities should be at least 30 to 70 seats at the optimum.

As the request density varies in Fig.~\ref{fig:SandCO2}, we see that areas with more requests require more buses to reach their \coo optimum. The maxima of both car and public transport switching probabilities increase as $\lambda_{tot}$ increases, resulting in greater absolute and relative \coo emission reductions: where 28.03\% can be saved at the maximum $\lambda_{tot}=10^5$ compared to the same system without ride-pooling ($B=0$), 42.40\% can be saved at $\lambda_{tot}=5 \times 10^5$. In the Base Case, 39.16\% can be reduced in comparison to a system without ride-pooling. This is not only because more people are persuaded to switch, but each vehicle carries more passengers, further increasing the efficiency of the system.

\begin{figure}[t]
\centering\includegraphics[width= \columnwidth]{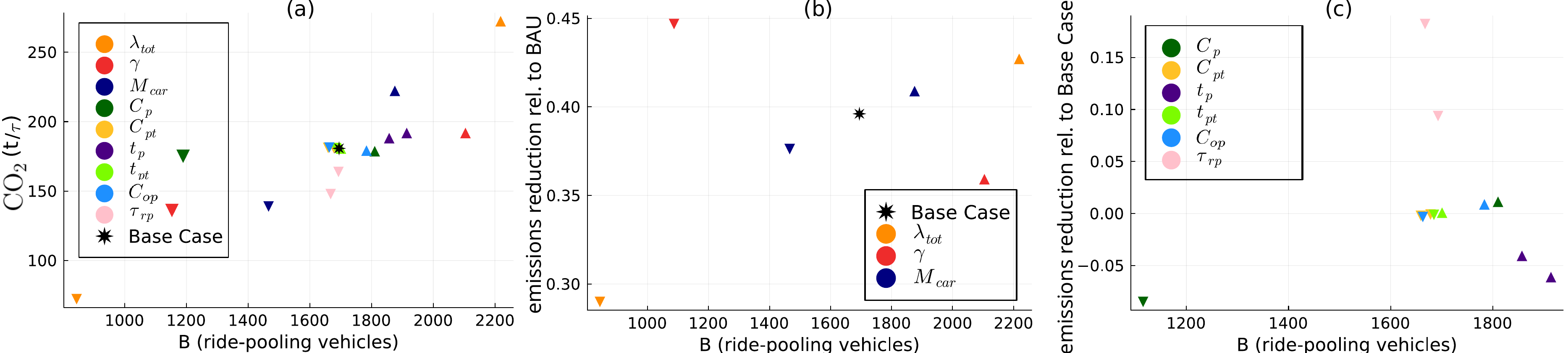}
\caption{Absolute and relative \coo in tons per time $\tau$ at optimal number of ride-pooling vehicles. The triangles pointing upwards show the increase in the respective parameter, the triangles pointing downwards show the decrease in the variation (exact values can be found in Tab.~\ref{tab:value definitions}). Fig.~\ref{fig:policy_oneplot})(a) Absolute \coo emissions at \coo minima range from less than 100 to more than 250 t \coo per $\tau$ in the system. b) \coo emissions reduction at \coo minimum relative to the emissions of the same system at fleet size $B=0$ (BAU: Business As Usual, a system without ride-pooling) for variations of system parameters ($\lambda_{tot}$, $\gamma$ and $M_{car}$). c) \coo emissions at \coo minimum relative to the \coo minimum of the Base Case (Base Case: variables as in Tab.~\ref{tab:value definitions}) for different policy parameters.}
\label{fig:policy_oneplot}
\end{figure}

In the following, we analyze the system's response to parameter changes, ranging from largely unchangeable regional parameters (total demand $\lambda_{tot}$, topological factor $\gamma$, initial modal share $M_{car}$) to policy parameters (parking price, public transport price, ride-pooling speed, lack of parking spaces causing delays, car purchase and maintenance costs to be included). The results in terms of carbon dioxide emission reductions can be seen in Fig.~\ref{fig:policy_oneplot} and Tab.~\ref{tab:differentcoo_minima}.

Fig.~\ref{fig:policy_oneplot}(a) shows the optimal number of vehicles in terms of \coo emissions for a range of parameter scenarios shown in a scatter plot. The black star indicates the "Base Case" in the sense that it uses all parameters identical to Tab.~\ref{tab:value definitions}. All other markers have one parameter changed relative to the Base Case scenario (ceteris paribus), with triangles pointing down indicating the parameter value was decreased and triangles pointing up indicating the parameter value was increased. The color of the marker indicates the changed parameter.

\begin{table}[b]
 \begin{center}
 \caption{The relative emission reductions are shown in Fig.~\ref{fig:policy_oneplot}(b) and (c). The system parameter modification on the one hand compares the \coo minimum with the \coo value of the same system for a ride-pooling fleet size of $B=0$. The policy parameter change on the other hand compares the \coo minimum with the modified parameter with the \coo minimum of the Base Case (with values as in Tab.~\ref{tab:value definitions}). The term BAU (Business as Usual) describes a system where no ride-pooling is introduced (B=0). Base Case, on the other hand, depicts the system with ride-pooling ($B>0$) and the parameter values as in Table I.}
    \label{tab:differentcoo_minima}
\begin{tabular}{lr}
    \begin{tabular}{l|l}
     & \textbf{$1-\frac{\coo^{\text{min}}}{\coo^{\text{B=0}}}$}\\
     \textbf{system parameter change}&rel. to BAU\\
    \hline
      $\lambda_{tot} = 10^5$ & $0.2803$ \\
      $\lambda_{tot} = 5 \times 10^5$ & $0.424$\\
    \hline
      $\gamma = 996$ & $0.355$  \\
      $\gamma = 213$ & $0.4417$  \\
    \hline
      $M_{car} = 0.28$ & $0.3704$  \\
      $M_{car} = 0.48$ & $0.4052$  \\
    \end{tabular}
    \hspace{4em}
    &

    \begin{tabular}{l|l}
      & \textbf{$1-\frac{\coo^{\text{min1}}}{\coo^{\text{min2}}}$}\\
       \textbf{policy parameter change}&rel. to Base Case\\
    \hline
      $C_p = 0$ \euro & $-0.0833$  \\
      $C_p = 15$ \euro & $0.0111$ \\
    \hline
      $C_{pt} = 1 $ \euro  & $-0.001$  \\
      $C_{pt}= 0.5 $ \euro& $-0.0021$  \\
    \hline
      $C_{op} = 0 $ \euro  & $-0.0033$  \\
      $C_{op}= 0.55 $ \euro& $0.0087$  \\
    \hline
      $\tau_{rp} = 0.4$ h & $0.1787$  \\
      $\tau_{rp} = 0.5$ h & $0.0919$  \\
    \hline
      $t_p = 0.5$ h & $-0.0401$  \\
      $t_p = 0.75$ h& $-0.0602$  \\
    \hline
      $\epsilon  = 1.4$ h & $-0.0012$  \\
      $\epsilon = 2.6$ h& $0.0007$  \\
    \end{tabular}
    \end{tabular}
\end{center}
\end{table}

In this calculation, system and policy parameters were considered separately. This is due to the fact that policy parameters can be adjusted by a government over a short term period, whereas system parameters are hardly adjustable. Rather, system parameters are considered as given depending on location and situation. Consequently, the calculation of relative values is also performed separately. Thus, system parameters evaluate the \coo savings in comparison to a system that does not incorporate ride-pooling. In contrast, policy parameters contrast the \coo minimum with the \coo minimum in the absence of parameter modification (Base Case).

Changes made to system parameters are shown in Fig.~\ref{fig:policy_oneplot}(b) and Tab.~\ref{tab:differentcoo_minima} (left column). They show the maximal emission reductions with ride-pooling, relative to the same system at fleet size $B$ equal to zero (BAU). The variation results show that the emission reduction for different demand $\lambda_{tot}$ can vary between 28.03\% for lower demand and 42.40\% for higher demand. This confirms the common knowledge that systems with higher demand are more suitable for the pooling of trips. The results demonstrate that varying the topological factor $\gamma$, to take values that have the topology of Ostfriesland ($\gamma = 213$) or Göttingen ($\gamma= 996$) leads to a range of 44.17\% to 35.5\% in \coo savings. This indicates that ride-pooling can be beneficial in simpler topologies, such as rural areas. The initial share of car drivers, denoted by $M_{car}$, shows that if the initial proportion of car usage within the system is higher, a relatively greater amount of \coo can be saved due to higher initial emissions.

Fig.~\ref{fig:policy_oneplot}(c) and Tab.~\ref{tab:differentcoo_minima} (right column) demonstrate the alterations made to the policy parameters. Here the \coo minimum with the changed parameter is compared to the \coo minimum with the Base Case parameters from Tab.~\ref{tab:value definitions}. It is interesting to see that even though some parameters are varied very much (e.g. $C_p$ from 6 \euro \space to 15 \euro), the change in the \coo minimum varies by only a few percentage points (in this case the \coo minimum decreases by only 1.11\%). Thus, we only look at the most interesting policy parameters that induce changes in \coo emission: ride-pooling travel time $\tau_{rp}$ produces most changes in emissions, as reducing travel time (e.g., by using express lanes or introducing high occupancy vehicle (HOV) lanes) can reduce emissions by 17.87\% for $\tau_{rp} = 0.4$ h. At the same time, emissions would increase by 8.33\% if the parking cost were reduced to $C_p = 0$ \euro, which shows that travel times and direct pricing, such as parking costs, can have a positive impact on emissions.

\subsection{Influence of the Switching Probability Function on Mode Choice}

While estimates of parameters can be found in the literature for various regions and policies, we are not aware of any published work on the functional form of the switching probability. However, the economic literature on similar mode choices (i.e., mode choice between private vehicle and public transport) most often uses logit functions \cite{quinet2004principles} by assigning a monetary cost to time.

\begin{table}[b]
\footnotesize
\begin{center}
\begin{tabular}{l|l|l|l|l|l|l|l}
    \textbf{abbrev.} &
    \textbf{switching probability function $P_{ref}$} &
    \textbf{$B_{opt}$} &
    \textbf{\coo($B_{opt}$)} &
    \textbf{$B_{10}$} &
    \textbf{\coo($B_{10}$)} &
    \textbf{$B_{5}$}&
    \textbf{\coo ($B_{5}$)}\\
    \hline
    original &
    $\frac{1}{\frac{t_s}{t_{ref}}+1}\frac{1}{\frac{C_B}{C_{ref}}+1}$&
    1695&
    39.16\%&
    5700&
    29.35\%&
    10500&
    12.77\%\\
    sum &
    $0.5\,\frac{1}{\frac{t_s}{t_{ref}}+1}+0.5\,\frac{1}{\frac{C_B}{C_{ref}}+1}$&
    1112  &
    65.07\%&
    8300&
    43.32\%&
    15700&
    17.51\%\\
    square&
    $\frac{1}{\frac{t_s}{t_{ref}}^2+1}\frac{1}{\frac{C_B}{C_{ref}}^2 +1}$ &
    2765 &
    39.27\%&
    7000&
    31.41\%&
    13700 &
    10.71\%\\
    logit &
   $\frac{1}{1+exp^{\beta*(f_{rp}-f_{ref})}}$ with $f = p+ht$&
    324 &
    91.59\%&
    10400&
    61.38\%&
    20400 &
    29.31\%\\
    upper b. &
    $\alpha+(-\alpha+\beta_2)\theta(-1+\frac{t_s}{t_{ref}}) \theta(-1+\frac{C_B}{C_{ref}})$&
    1 &
    96.12\% &
    11400&
    62.83\%&
    22800 &
    29.54\%\\
    lower b. &
    $\beta_1 \theta(1-\frac{t_s}{t_{ref}}) \theta(1-\frac{C_B}{C_{ref}})+(\alpha-\beta_1)\theta(\frac{t_s}{t_{ref}}) \theta(\frac{C_B}{C_{ref}})$&
    2956 &
    15.40\%&
    2956&
    15.40\%&
    5700 &
    7.39\% \\
\end{tabular}
\end{center}

\caption{Functional forms of the switching probability functions $P_{ref}$ and the resulting relative \coo emission reduction of the optimal fleet size $B_{opt}$ compared to a system without ride-pooling (B=0). The same is calculated for a midibus fleet size $B_{10}$ and a minibus fleet size $B_{5}$.} The parameters ($f_{rp}$ for ride-pooling and $f_{ref}$ for the former mode of transport) in the logit function are generalized to include not only the financial cost $p$, but also the travel time $t$ multiplied by the financial value of time $h = 0.5$, which is estimated as a fixed system parameter ($f = p+ht$).
The upper and lower bounds are the functions of the highest and lowest values $P_{ref}(t_s,C_B)$ for all parameter values. They are constructed as combinations of heaviside step functions $\theta(.,.)$. More visualizations of all switching probability functions are given in the Supplemental Information.
\label{tab.sr}
\end{table}

\begin{figure}[t]
    \centering
    \includegraphics[width=0.9\textwidth]{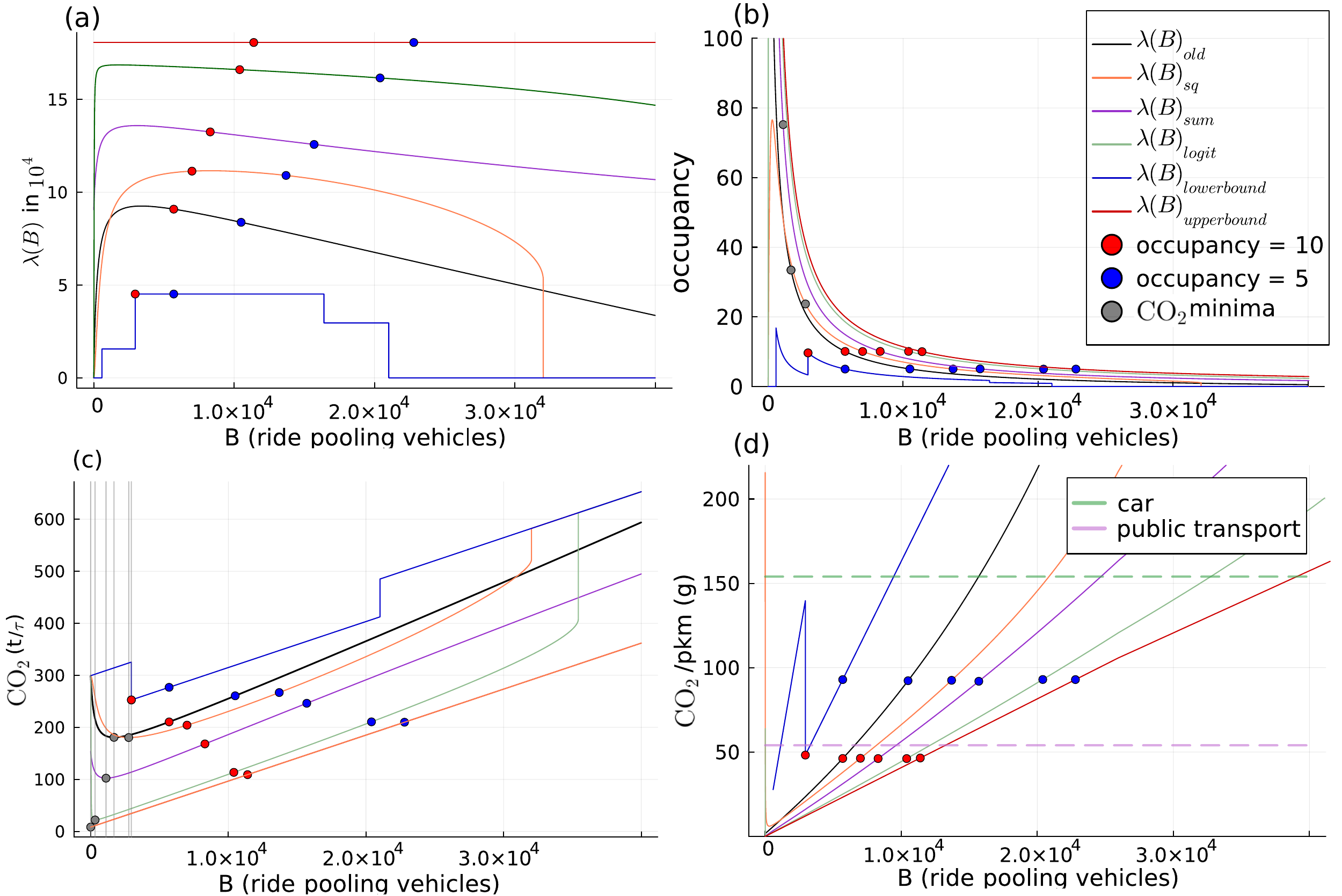}
    \caption{Various functional forms for the switching probability function $P_{ref}$ lead to qualitatively similar results a) Ride-pooling users over ride-pooling fleet size $B$ for different switching probability functions. b) Average occupancy of the ride-pooling vehicles for various $P_{ref}$. c) \coo emissions over fleet size for various $P_{ref}$. d) \coo emissions per passenger kilometer are similar to public transport (dotted pink line) at occupancies of 10 and still far below personal cars (dotted green line) at occupancies of 5.}  
    \label{fig.alternative_sr}
\end{figure}

Here we want to explore the space of all functions that are compatible with the switching probability constraints i) through iv). Thus, we define the upper bound switching probability function as the function compatible with the constraints with the highest values $P_{ref}(t_s,C_B)$ for any pair $(t_s,C_B)$, and the lower bound switching probability as the function compatible with the constraints with the lowest values $P_{ref}(t_s,C_B)$ for any pair $(t_s,C_B)$.
The upper and lower bound functions are therefore combinations of step functions, their functional forms are given in Tab.~\ref{tab.sr}. Note that these functional forms ignore behavior at infinity (upper bound) or zero (lower bound), since we are only analyzing the system at finite times and costs.
The switching probability functions, $P_{ref}$, presented in Tab.~\ref{tab.sr} were numerically evaluated as the consistency equation Eq.~\ref{eq.sr} has no closed functional form. Their shapes are shown in Fig. 1 of the Supplemental Information (SI).
The resulting curves for ride-pooling requests $\lambda(B)$, occupancy $O$, and \coo emissions per $\tau$ and per person-kilometer are shown in Fig.~\ref{fig.alternative_sr}.

Fig.~\ref{fig.alternative_sr}(a) shows that the general behavior of the ride-pooling switching probability is independent of the exact functional form, but rather responds similarly to changes in the ride-pooling fleet size B. Across all sampled switching probability functions $P_{ref}$, we observe first an increase and then a decrease in the number of ride-pooling users as B increases. The occupancy (Fig.~\ref{fig.alternative_sr} (b)) decreases with the fleet size but in case of the lower bound function this decrease in not monotonous. The occupancies at the \coo minimum vary drastically between different switching probability functions. We therefore also analyze equal-occupancy fleet sizes $B_{10}$ and $B_5$, which are the (maximum) fleet sizes of average occupancy 10 and 5 respectively. 

The system-wide \coo emissions per $\tau$, calculated according to Eq.~\ref{eq.coo} using the parameter values from Tab.~\ref{tab:value definitions} and shown in Fig.~\ref{fig.alternative_sr}(c) first decrease with increasing B as ride-pooling is introduced, and increase again when the fleet size becomes so large that the average occupancy per ride-pooling vehicle decreases. While the \coo emissions at the minimum range from 15\% reduction to 96\% reduction, this range becomes significantly narrower as vehicle sizes are constrained, resulting in 15\% - 63\% reduction for average occupancy of 10 and 7\% - 30\% reduction for an average occupancy of 5.

We can compare the \coo emissions of ride-pooling assuming any of the proposed switching probability functions in Fig.~\ref{fig.alternative_sr}(d), and compare it to the \coo emissions of other modes as reported in the literature \cite{UBA2020}. 
Average \coo emissions per passenger kilometer (pkm), denoted by $CO_2^{rp}{pkm}$ and given in grams per passenger kilometer (g/pkm) are calculated as the \coo emissions from ride-pooling divided by the number of users and the distance traveled (all per $\tau$):
\begin{equation}
    CO_2^{rp}{pkm}=\frac{(\frac{E_{rp}Bl}{\lambda(B)})}{l}=\frac{E_{rp} B}{\lambda(B)}
    \label{eq.coo_pc}
\end{equation}
We find that even with larger-than-optimal fleet sizes, pkm \coo emissions remain well below those of cars for occupancies of 5, and even below those of public transport for occupancies of 10 (Fig.~\ref{fig.alternative_sr}(d)).

\section{Discussion and Conclusion: Make ride-pooling time-efficient}

This work proposes a new framework for estimating the impact of introducing ride-pooling on carbon emissions. We analyzed the potential for reducing \coo emissions based on a) parameter variations, assuming a simple switching probability function compatible with the constraints i) - iv) set up in the methods section, and b) variations over the range of constraint-compatible switching probability functions, with a Berlin parameter set. We based our calculations on the collective dynamics of ride-pooling, derived in \cite{molkenthin2020scaling}, together with four assumptions about the switching behavior of customers. While the model is too simple to make precise quantitative conclusions, we are able to draw a number of interesting qualitative conclusions.

We find that, depending on the parameter set, the system-wide reduction potential based on the analytical switching probability function is between 28\% to 44\%, with policy parameters adding only another [$-8$, $+18$]\%  of variation. In particular, free parking can reduce the acceptability of ride-pooling, leading to a 8.3\% increase in \coo, but further increasing the parking fee has a diminishing return of only another 1.1\%. Reducing the travel time for ride-pooling (e.g., by creating priority lanes) can increase the acceptance of ride-pooling and thus further reduce \coo emissions by up to 18\%. 
While we have varied each parameter independently, combining several policies may further improve ride-pooling viability and sustainability.
Note that other policies, such as raising awareness of the true cost of private car use (changing $C_{op}$ in the model), changing the price of public transport (changing $C_{pt}$), or changing the speed of public transport (changing $\epsilon$) have negligible effects. In particular, this relative robustness to cost changes may be related to the fact that we assume cost sharing in Eq.~\ref{eq.Cb}, unlike previous studies which only assume a fixed 50\% discount \cite{de2023ride} on the taxi price. While dynamic pricing is difficult to implement, we expect that pricing based on average occupancy will produce the same results.

The \coo optima of all switching probability functions for the city of Berlin are below 3000 vehicles. The minimum \coo savings are between 15 and 96\%, so more information about realistic switching behavior would be needed to make accurate predictions. However, since the average occupancy in these cases is very high, they would require full-size buses as pooling vehicles, which may be unfeasible. Limiting the occupancies to 5 or 10 persons per ride-pooling vehicle models the use of small- to medium sized vehicles, often used to facilitate maneuvering on residential streets. To account for fluctuations in occupancy, vehicle sizes should be larger than the average occupancy. The results presented in \cite{zech2022collective} suggest that vehicles about 50\% larger than average occupancy would be able to handle the vast majority of requests.
We find that the fleet size required to serve Berlin ranges from 3,000 to 11,400 vehicles for an average occupancy of 10 (vehicle size approximately 15 to 20 seats) or 5,700 to 22,800 for an average occupancy of 5 (vehicle size approximately 7 to 10 seats). Although these ranges are not very narrow, they assume very little about customers' mode choice behavior. 
These results are in line with a recent simulation study \cite{schmaus2023shared} empirically gauged on origin-destination data, which found that about 26,000 6-seater ride-pooling vehicles would be needed to meet Berlin's entire car transport needs. 

We observe that the switch to ride-pooling by former public transport users has a seemingly paradoxical positive effect on \coo emissions. This is again due to occupancy-based pricing, as the presence of additional customers (even if they come from public transport) reduces the price per ride-pooling user and thus incentivizes car users to switch as well.

Further analysis is needed to increase the validity of results. This would make it directly applicable to real-world scenarios, including a transient introduction process, psychological factors that may inhibit ride-pooling acceptance, and agent-based, interacting decision processes as well as direct empirical observations. Most importantly, the effect of detours and stop times should be investigated comprehensively, for example using perturbation theory based on the perfect service limit. Accounting for friction due to stopping while loading passengers will likely require moving from analytically solvable models to numerical simulation models. The model should also be adapted to allow for different pricing schemes, such as capping the price so as not to undercut public transport. For such studies, this model can be helpful in guiding the selection of parameter ranges to minimize computational effort, or in providing a framework of observables for empirical analysis.

\section*{Conflict of interests}
The authors declare no competing interests.

\section*{Acknowledgements}
We thank Karolin Stiller and Alexander Schmaus for fruitful discussions. Felix Creutzig acknowledges funding by the European Union under grant agreement No. 101056810.

\bibliographystyle{IEEEtran}
\bibliography{ecobuslit}
\end{document}